\documentclass[intlimits,twoside,a4paper]{article}
\usepackage{amsmath,amssymb}
\usepackage{graphicx}
\usepackage{subfigure}

\usepackage[T2A]{fontenc}
\usepackage[cp1251]{inputenc}

\usepackage[eqsecnum]{cmpj2}
\issue{2013}{16}{1}{13602}

\doinumber{10.5488/CMP.16.13602}

\title[The application of a combined potential in fluoride glass by HRMC]%
{Study of the validity of a combined potential model using the hybrid reverse Monte Carlo method in fluoride glass system}

\author[S.M. Mesli \textsl{et al.}]{S.M.~Mesli\refaddr{label1,label2}\thanks{E-mail: msidim1975@gmail.com},
         M.~Habchi\refaddr{label1,label3}, M.~Kotbi\refaddr{label2}, H.~Xu\refaddr{label4}}

\addresses{
\addr{label1} Departement of physics, LPTPM, Hassiba Ben Bouali University, Chlef 02000, Algeria
\addr{label2} Department of Physics, LPM, A.B. Belkaid University, BP 119 Tlemcen 13000, Algeria
\addr{label3} Preparatory School on Sciences and Technics, BP 165 RP Bel Horizon, Tlemcen 13000, Algeria
\addr{label4} Institut of Physics, Paul Verlaine University, 57000 Metz, France
}

\date{Received April 25, 2012, in final form August 30, 2012}

\authorcopyright{S.M.~Mesli, M.~Habchi, M.~Kotbi, H.~Xu, 2013}

\begin{document}

\maketitle

\begin{abstract}
The choice of appropriate interaction models is among the major disadvantages of conventional methods such as molecular dynamics and Monte Carlo simulations. On the other hand, the so-called reverse Monte Carlo (RMC) method, based on experimental data, can be applied without any interatomic and/or intermolecular interactions. The RMC results are accompanied by  artificial satellite peaks. To remedy this problem, we use an extension of the RMC algorithm, which introduces an energy penalty term into the  acceptance criteria. This method is referred to as  the hybrid reverse Monte Carlo (HRMC) method. The idea of this paper is to test the validity of a combined potential model of coulomb and Lennard-Jones in a fluoride glass system BaMn$M$F$_{7}$ ($M={}$Fe,V) using HRMC method. The results show a good agreement between experimental and calculated characteristics, as well as a meaningful improvement in partial pair distribution functions. We suggest that this model should be used in calculating the  structural properties and in describing the  average correlations between components of fluoride glass or a similar system. We also suggest that HRMC could be useful as a tool for testing the interaction potential models, as well as for conventional applications.
\keywords RMC simulation, unrealistic features, hybrid RMC simulation, Lennard-Jones potential, fluoride glass
\pacs 61.20.Ja,31.15.P,05.45.Pq
\end{abstract}

\section{Introduction}

Several simulation methods have been applied to the study of different types of ordered and disordered system (liquids, glass, polymers, crystals and magnetic materials)  \cite {1,2,3}. For example, molecular dynamics (MD) and Monte Carlo (MC) simulations are frequently used to investigate the physical phenomena that are not easily accessible via experiment  \cite{4}. The fundamental input to such simulations is a potential model, which can describe the interactions between atoms or molecules and calculate the energy of the system  \cite{5}. The development of an adequate potential for a classical application poses a major problem.

On the other hand, another method of modelling called a reverse Monte Carlo (RMC), based on the experimental data, has an advantage of being  applied without any interaction potential model  \cite{6}. This method completes the experiment by computing the pair distribution functions (PDFs) between each two different components of a system. The RMC simulation results still display some physically unrealistic aspects, such as the appearance of artifacts in PDFs. This can be due to a limited set of experimental data and/or due to the non-unique RMC models  \cite{7}. In order to solve this problem, we apply a modified simulation protocol based on RMC algorithm, whose physical or chemical constraint  is introduced based on the understanding of the material being modelled  \cite{8}. Thus, we refer to it as  the hybrid reverse Monte Carlo (HRMC). The latter combines the features of MC and RMC methods  \cite{9}.

In the present work, two types of potentials, which take the long and short range interaction into account, are combined to build the interaction potential model, namely \textit{Coulomb plus Lennard-Jones model}. This potential can be applied at the atomic scale, and at several states of materials. In this sense, we have chosen it as a physical constraint applied at a glassy
state.

Fluoride glass BaMn$M$F$_{7}$ ($M={}$Fe,V, assuming isomorphous replacement) is a system in which  our energetic constraint is applied. The neutron data for both BaMnFeF$_{7}$ and BaMnVF$_{7}$ glasses were recorded at ILL (Grenoble)  \cite{10}. The estimated isomorphous replacement between Fe$^{3+}$ and V$^{3+}$ is well supported by the crystal chemistry in fluoride compounds. This substitution in fluoride materials appears to be  quite interesting for carrying out a neutron scattering experiment  \cite{10}.

The purpose of this article is to test the effect and to investigate the validity of the added potential using the HRMC simulation \cite{11}, with the aim of adapting this potential to fluoride glasses or similar system studied by conventional methods.

\section{Simulation details}
\label{sim-det}

\subsection{Reverse Monte Carlo method}
\label{rmc-mt}
Since the RMC method has already been  described in detail  \cite{12,13,14}, we will only give its brief summary. Its aim is to construct large, three-dimensional structural models that are consistent with total scattering structure factors $S(Q)$  obtained from diffraction experiments within fixed standard deviation.

\looseness=-1A modification of the Metropolis Monte Carlo (MMC) method is used in  \cite{15}. Instead of minimizing the potential term like in the classical methods of MD and MC, the difference $\chi^{2}$ between the calculated and the experimental partial distribution functions $G(r_{i})$ is the quantity to be minimized via random movements of particles. The partial distribution function $G(r_{i})$ is written as:
\begin{equation}
\label{Gder-def}
G^{\mathrm{RMC}}(r) = \frac{N_{\mathrm{RMC}}(r)}{4\pi r^{2}\rho \varDelta r}\,,
\end{equation}
where $ \rho $ is the atomic number density and $N_{\mathrm{RMC}}(r)$ is the number of atoms at a distance between $r$ and $r+\rd r$ from a central atom, averaged over all atoms as centers.

$G^{\mathrm{RMC}}(r)$ is the inverse Fourier transform of the structure factor $S^{\mathrm{RMC}}(Q)$ depending on the wave vector $Q$ and expressed as:
\begin{equation}
\label{SdeQ-def}
S^{\mathrm{RMC}}(Q) = 1+\rho \int_{0}^{\infty} 4\pi r^{2}\left[G^{\mathrm{RMC}}(r)-1\right] \frac{\sin  Qr}{Qr}\rd r.
\end{equation}
The quantity to be minimized is written as:
\begin{equation}
\label{Xi2-def}
\chi^{2}=\sum_{i}\left\{\frac{\left[G^{\mathrm{RMC}}(r_{i})-G^{\mathrm{EXP}}(r_{i})\right]^{2}}{\wp^{2}(r_{i})}\right\}\,.
\end{equation}
For any couples of partial distribution functions, $G^{\mathrm{RMC}}(r_{i})$ is obtained by RMC configurations, and \linebreak $G^{\mathrm{EXP}}(r_{i})$ is the experimental result. $\chi^{2}$ is calculated by using a statistical measure error estimated by a standard deviation $\wp(r_{i})$ which is supposed to be uniform and independent of distance $r_{i}$.
RMC simulation starts with an appropriate initial configuration of atoms. When modelling crystalline materials, this configuration will have atoms in their average crystallographic positions, and will contain several unit cells. While modelling the non-crystalline materials, an initial algorithm will be required to generate a random distribution of atoms without unreasonably short inter-atomic distances. Atoms are selected and moved randomly to obtain a new configuration after each move. The $G(r_{i})$ of a new configuration as well as the $\chi^{2}$ are calculated. If $\chi^{2}_{\mathrm{new}}$ is less than $\chi^{2}_{\mathrm{old}}$, the agreement between the experimental and the current configuration is improved by a move. Thus, the move is accepted and another move is made. If $\chi^{2}_{\mathrm{new}}$ increases, it is not rejected outright but accepted with a probability $P_{\mathrm{acc}}$ given by:
\begin{equation}
\label{Pacc-def}
P_{\mathrm{acc}}=\mathrm{exp}\left(-\frac{\chi_{\mathrm{new}}^{2}-\chi_{\mathrm{old}}^{2}}{2}\right).
\end{equation}
The process is then repeated until $\chi^{2}$ fluctuates around an equilibrium value.

The resulting configuration should be a three-dimensional structure compatible with the experimental partial function. Simulation parameters such as the number of atoms, the density, and the length of the simulation box are given in table~\ref{tbl-smp1}. The cut-offs (geometric constraint) between pairs of atoms are presented in table~\ref{tbl-smp2}.
\begin{table}[h]
\caption{$N_{i}$ indicates the number of atoms of species $i$ ($i={}$ Ba, Fe, Mn, F), $\rho$ is the total atomic density and $L$ is the length of the simulation box.}
\label{tbl-smp1}
\vspace{2ex}
\begin{center}
\begin{tabular}{|c|c|c|c|c|c|c|}
\hline\hline
$N_{\mathrm{Ba}}$&$N_{\mathrm{Fe}}$&$N_{\mathrm{Mn}}$&$N_{\mathrm{F}}$&$N$&$ \rho $&$L$\strut\\
\hline
\hline
$500$&$500$&$500$&$3500$&$5000$&$0.0710$&$20.647$\strut\\
%\cline{2-10}
\hline\hline
\end{tabular}
\end{center}
\end{table}

When satisfactory agreement between experimental and theoretical data sets is obtained, detailed structural data such as coordination number, bond angle distribution functions, and PDFs can be calculated from atomic networks, being averaged over many MC configurations that are consistent with the experimental data.
\begin{table}[h]
\caption{$S_{ij}$ cut-offs between atomic pairs.}
\label{tbl-smp2}
%\vspace{1ex}
\begin{center}
\begin{tabular}{|c|c|c|c|c|c|c|c|c|c|c|}
\hline\hline
atomic pairs&BaBa&BaFe&BaMn&BaF&FeFe&FeMn&FeF&MnMn&MnF&FF\strut\\
\hline
\hline
$S_{ij}$(\AA)&$3.50$&$3.00$&$3.00$&$2.00$&$2.70$&$2.70$&$0.50$&$2.65$&$1.20$&$0.40$\strut\\
%\cline{2-10}
\hline\hline
\end{tabular}
\end{center}
\end{table}

\vspace{-5mm}
\subsection{Hybrid reverse Monte Carlo method}

The RMC method may be used in studying the disordered crystalline materials at an atomistic level  \cite{16}. The lack of a potential has the disadvantage of RMC models having no thermodynamic consistency  \cite{12}. The RMC simulation results still display some artificial satellite peaks at the level of PDFs. This can be due to the set of experimental data restricted to only total  distribution functions and/or to the nonuniqueness problem. To remedy this problem, we refer to it as the HRMC simulation \cite{2,9,11,12,17,18}.
In the HRMC method, we introduce an energy constraint as a combined potential in addition to the common geometrical constraints derived from the  experimental data. The agreement factor $\chi^{2}$ becomes:
\begin{equation}
\label{Xi2HRMC-def}
\chi^{2}=\sum_{i}\left\{\frac{\left[G^{\mathrm{RMC}}(r_{i})-G^{\mathrm{EXP}}(r_{i})\right]^{2}}{\wp^{2}(r_{i})}\right\}+\frac{\omega U}{k_{\mathrm{B}}T}\,.
\end{equation}
Herein, $ U $ denotes the total potential energy. $\omega$ is a weighting parameter. In our RMC code, $\omega$ is between 0 and 1.
Thus, the acceptance criteria expressed by the conditional probability is now given as:
\begin{equation}
\label{PaccHRMC-def}
P_{\mathrm{acc}}=\mathrm{exp}\left[-\frac{\left(\chi_{\mathrm{new}}^{2}-\chi_{\mathrm{old}}^{2}\right)}{2}\right]\mathrm{exp}\left(-\frac{\Delta U}{k_{\mathrm{B}}T}\right).
\end{equation}
It is assumed that  $ \Delta U=U_{\mathrm{new}}-U_{\mathrm{old}} $ is the energy penalty term, $ U_{\mathrm{new}} $ and $ U_{\mathrm{old}} $ are the energies of the new and old configurations, respectively.

The potential energy function between the $i^{\mathrm{th}}$ and the $j^{\mathrm{th}}$ particles takes the following general form:
\begin{equation}
\label{U-def}
U_{ij}=\dfrac{q_{i}q_{j}}{4\pi \varepsilon_{0}r_{ij}}+4\varepsilon\left[\left(\dfrac{\sigma}{r_{ij}}\right)^{12}-\left(\dfrac{\sigma}{r_{ij}}\right)^{6}\right],
\end{equation}
where $q_{i}$ and $q_{j}$ are the charges of individual ions $i$ and $j$,  $r_{ij}$ is the atomic distance, $\varepsilon_{0}$ is the permittivity of free space, $\varepsilon$ is a parameter characterizing the depth of the potential well and $\sigma$ is the minimal distance between the interacting particles at which the potential of Lennard-Jones is zero  \cite{19}.

The interaction potential $U_{ij}$ is composed of two terms. The first one is the Coulomb interaction potential which takes the long-range interactions into account  \cite{20}. The second term is the Lennard-Jones potential. It takes the short-range interactions into account. In statistical physics, the Lennard-Jones potential is frequently used to model liquids and glasses. Our combined potential can be used at the atomic scale, and at several states of materials. Note that a disordered system at a glassy state shows a better structural organization compared to the liquid state. Hence,  we propose to apply  this potential in a fluoride glass system.

First, we should determine the potential parameters. Concerning Coulomb potential parameters, we only need charge fractions of all atomic species. On the other hand, the expressions of $\varepsilon$ and $\sigma$ of Lennard-Jones potential parameters for the similar interactions such as barium-barium, manganese-manganese, iron-iron, and fluor-fluor, are presented in table~\ref{tbl-smp3}. As concerns the eight remaining interactions, we try to use the Lorentz-Berthelot mixing rule equation as follows  \cite{21}:
\begin{equation}
\label{SigAB-def}
\sigma_{AB}=\dfrac{(\sigma_{AA}+\sigma_{BB})}{2}\,,
\end{equation}
\begin{equation}
\label{EpsAB-def}
\varepsilon_{AB}=\sqrt{\varepsilon_{AA}\varepsilon_{BB}}\,.
\end{equation}

\begin{table}[h]
\caption{Lennard-Jones potential parameters for similar interactions.}
\label{tbl-smp3}
%\vspace{1ex}
\begin{center}
\begin{tabular}{|c|c|c|}
\hline\hline
Pair functions&$\vphantom{\dfrac{\varepsilon'}{k'_{\mathrm{B}}}}\dfrac{\varepsilon}{k_{\mathrm{B}}}[K]$&$\sigma [\text{\AA{}}]$ \strut\\
\hline
\hline
Ba--Ba&$226.30$&$3.820$~\cite{22}\strut\\
\hline
Fe--Fe&$6026.70$&$2.319$~\cite{23}\strut\\
\hline
Mn--Mn&$5907.90$&$2.328$~\cite{23}\strut\\
\hline
F--F&$52.80$&$2.830$~\cite{24}\strut\\
\hline\hline
\end{tabular}
\end{center}
\end{table}

The RMC modelling of BaMn(Fe,V)F$_{7}$ is taken based on the two experimental structure functions of BaMnFeF$_{7}$ and BaMnVF$_{7}$ glasses obtained using the neutrons scattering technique. The study of the validity of our potential model is based upon comparing the experimental and simulation results.

\section{Results and discussions}
\label{res-disc}
\subsection{Total correlation functions}
\label{TC-funct}

\begin{figure} [h]
\centerline{
\includegraphics[width=7cm]{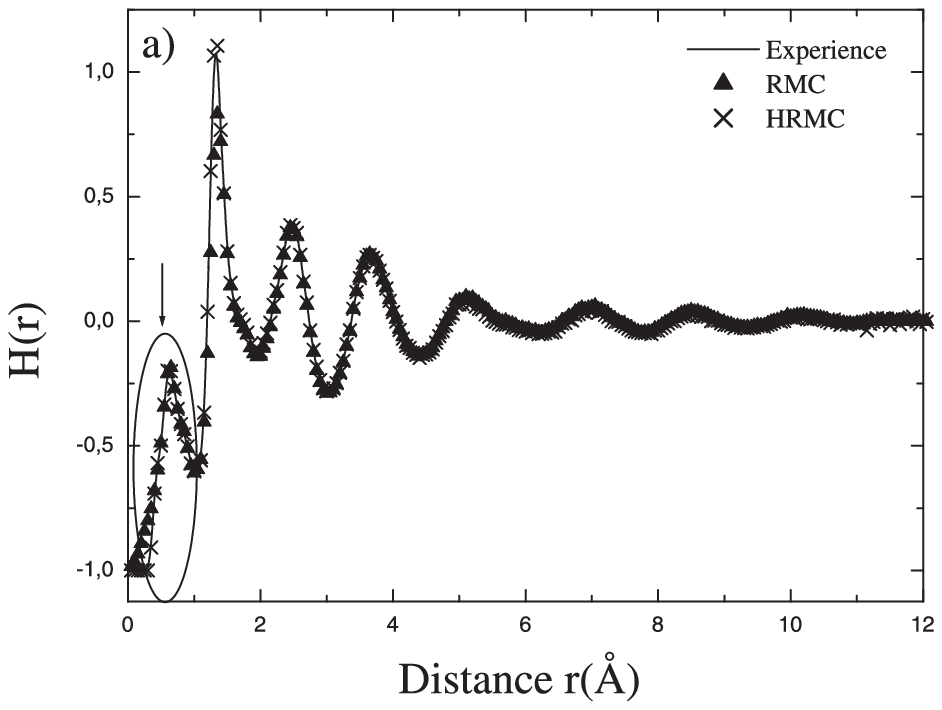}\hfill
\includegraphics[width=7cm]{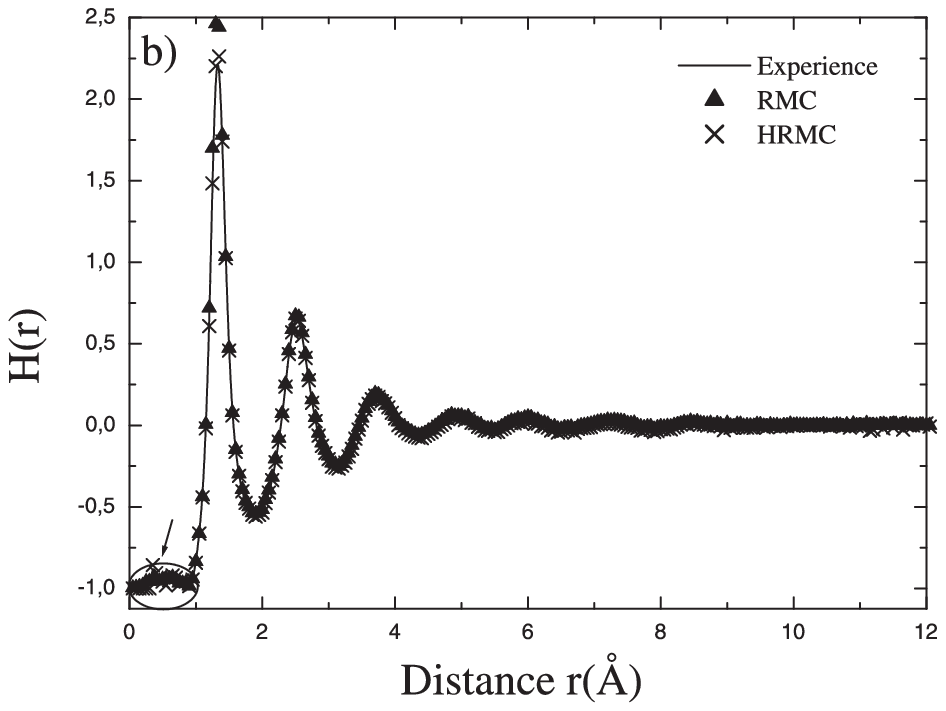}}
\caption{HRMC, RMC and experimental data: total distribution function $G(r)$ of (a) BaMnFeF$_{7}$ and (b) BaMnVF$_{7}$ at the glassy state; total correlation functions are represented $H(r)=G(r)-1$.}
\label{fig1}
\end{figure}

A comparison with experimental data is of primary importance in order to validate the  results of computer simulation methods that use interaction potential models  \cite{25}. In this work, the RMC code takes into account $G(r)$ the inverse Fourier transform of the total scattering structure factors $S(Q)$. Thus, the quantity used for this purpose is the total distribution function. Note that it is easy to use total correlation functions equivalent to total distribution functions: $H(r)=G(r)-1$  \cite{11,26}. Figure~\ref{fig1}, provides the total correlation functions for two structures of glassy states, i.e., BaMnFeF$_{7}$ in figure~\ref{fig1}~(a) and BaMnVF$_{7}$ in figure~\ref{fig1}~(b), and this is compared to the experimental results  for RMC and HRMC methods.

We notice that the fluoride glass has preserved its stability after incorporation of interaction potential, and the system is not disturbed. The results of total correlation functions calculated by RMC and HRMC are in excellent agreement with the results  obtained by experimental neutron diffraction. Thus, there is no conflict between the used method and the applied constraint. In fact, we can deduce that the present potential is valid. In order to better highlight this validity, curves of PDFs will be displayed \cite{9,11,25,26,27}.

\subsection{Pair distribution functions}
\label{Pd-funct}

\begin{figure} [!h]
\centerline{
\includegraphics[width=7cm]{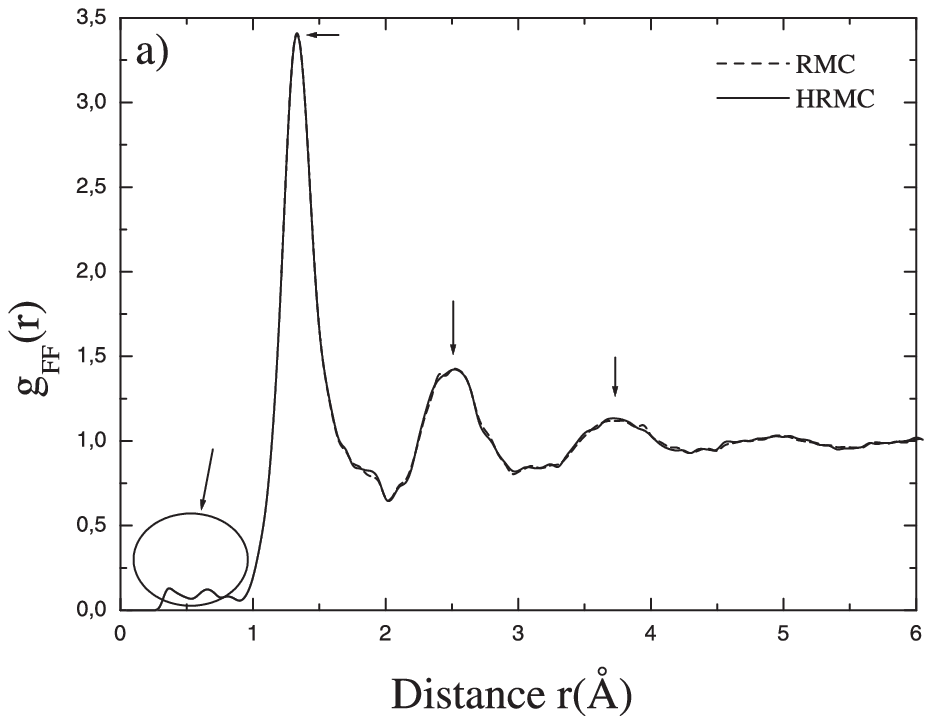}\hfill
\includegraphics[width=7cm]{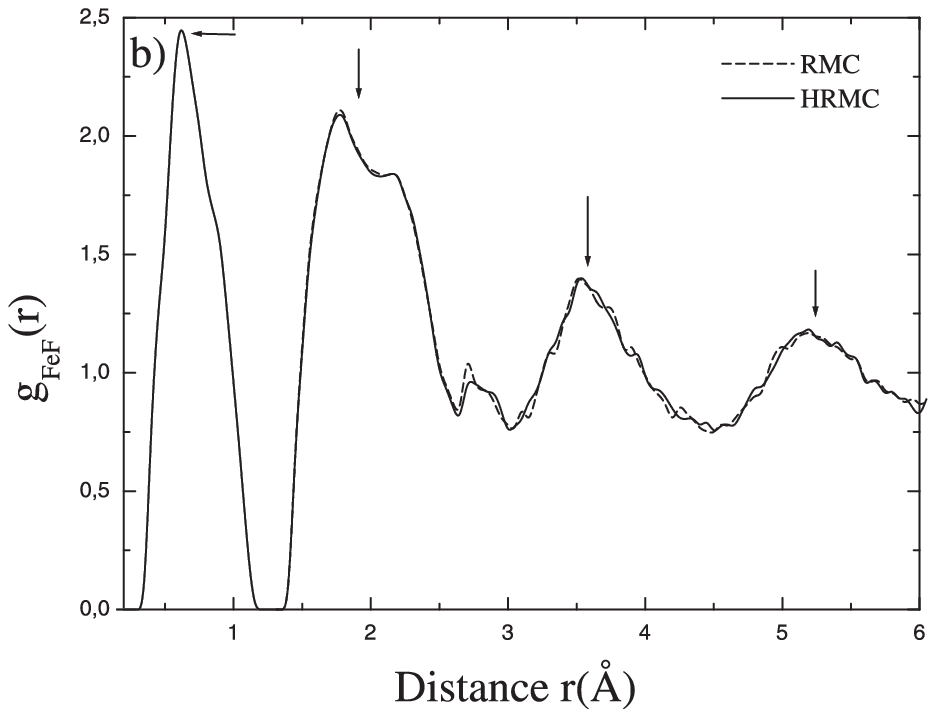}}
\centerline{
\includegraphics[width=7cm]{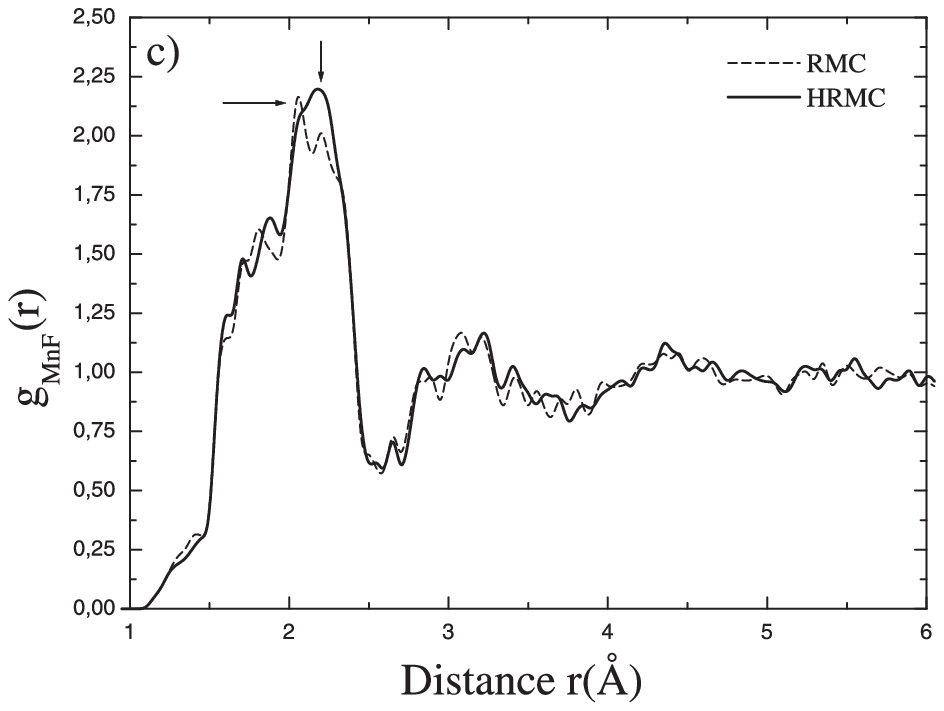}\hfill
\includegraphics[width=7cm]{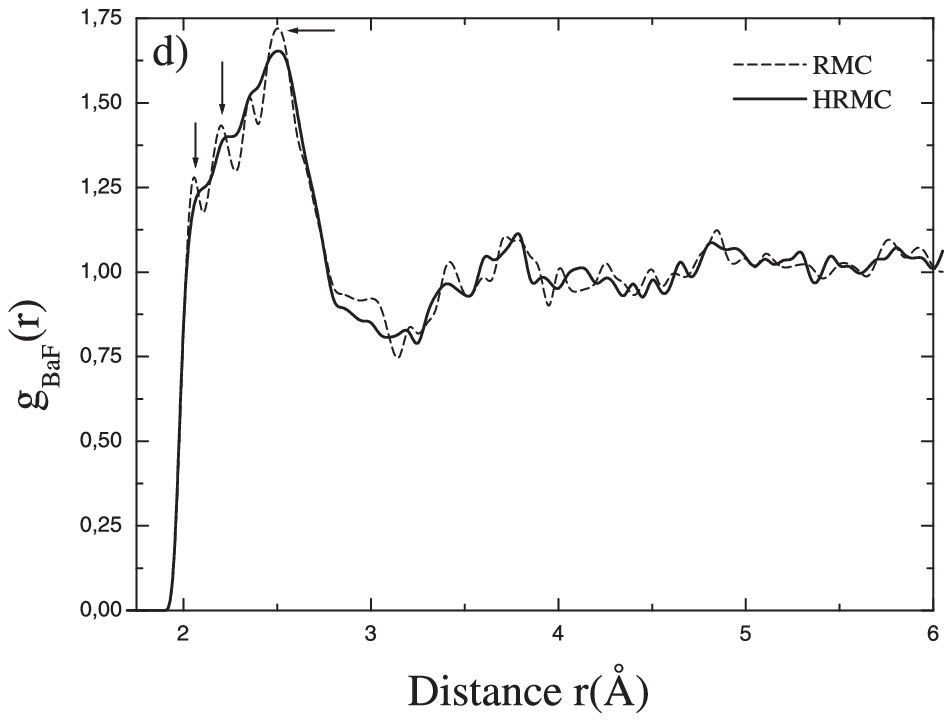}}
\caption{RMC and HRMC: partial pair distribution functions of $g_{\mathrm{FF}}(r)$ (a), $g_{\mathrm{FeF}}(r)$ (b), $g_{\mathrm{MnF}}(r)$ (c) and $g_{\mathrm{BaF}}(r)$ (d). }
\label{fig2}
\end{figure}

One could start with the fluoride interactions; figure~\ref{fig2} provides the PDFs for individual atomic species, namely $g_{\mathrm{FF}}(r)$ [figure~\ref{fig2}~(a)], $g_{\mathrm{FeF}}(r)$ [figure~\ref{fig2}~(b)], $g_{\mathrm{BaF}}(r)$ [figure~\ref{fig2}~(c)] and $g_{\mathrm{MnF}}(r)$ [figure~\ref{fig2}~(d)] calculated  by RMC and HRMC simulation. The fit of the obtained RMC results is effected by the value of the weighting parameter $\omega$. We choose the weighting factor to give the minimum fit while yielding structures that are physically realistic. Herein, we have chosen ($\omega = 20 \% $).

\newpage
The results obtained for PDFs, show a better accord between the used method and the applied potential, and this appears clearly at the level of $g_{\mathrm{FF}}(r)$ and $g_{\mathrm{FeF}}(r)$. The coordinations are reproduced well in each distribution [see arrows in figure~\ref{fig2}~(a)  and (b)]. This is a consequence of the number of fluor particles which greatly  exceeds the others constituting the fluoride glass.
 The (O--H) bonding distance in a water molecule is smaller than 1~\AA{}  \cite{11, 28}. The characteristics  below 1~\AA{} [see circle in figure~\ref{fig1}~(a) and (b)] are artifacts, resulting from Fourier errors, while transforming the measured data into  $G(r)$, as well as due to a great number of fluor particles; as seen by circle in figure~\ref{fig2}~(a); it is the only distribution which comprises an artifact, at the distance below 1~\AA{}.
Note that the direct application of RMC, developed PDFs accompanied by artifacts. The latter  appear at the level of $g_{\mathrm{MnF}}(r)$ and $g_{\mathrm{BaF}}(r)$ [see arrows in figures~\ref{fig2}~(c) and (d)].
Upon comparing with the results obtained by the HRMC simulation, a good smoothing is observed.

\begin{figure}[!h]
\centerline{
\includegraphics[width=7cm]{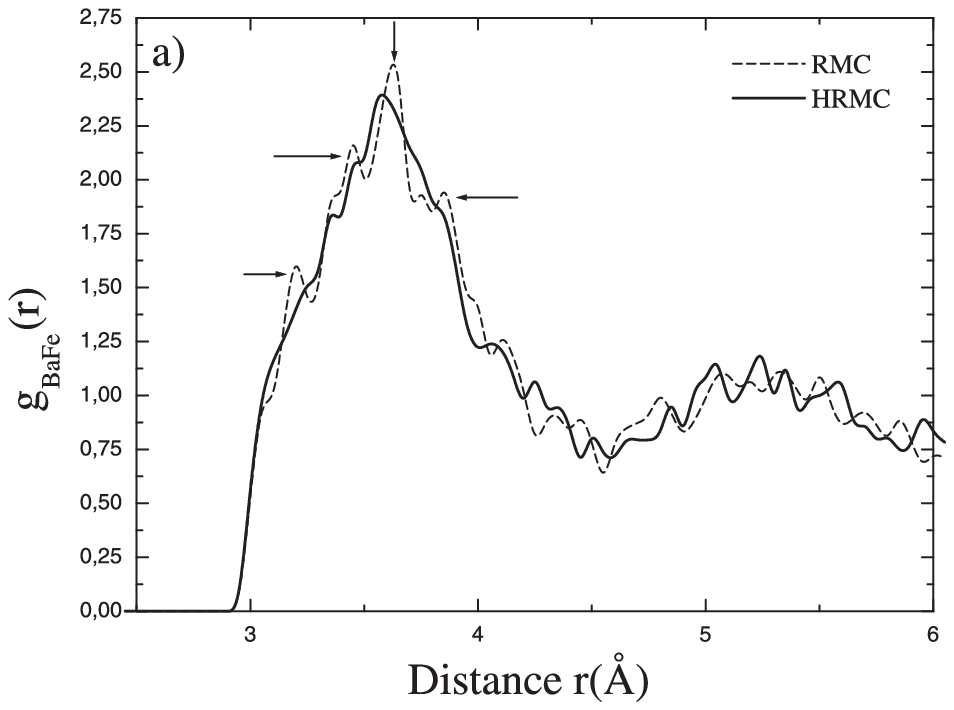}\hfill
\includegraphics[width=7cm]{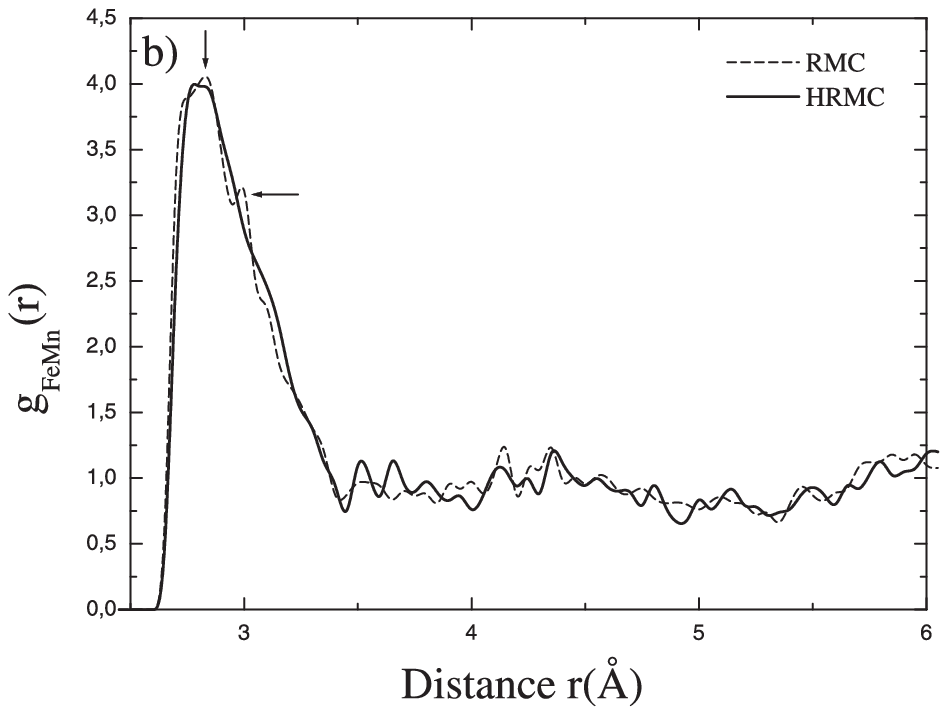}}
\vspace{3mm}
\centerline{\includegraphics[width=7cm]{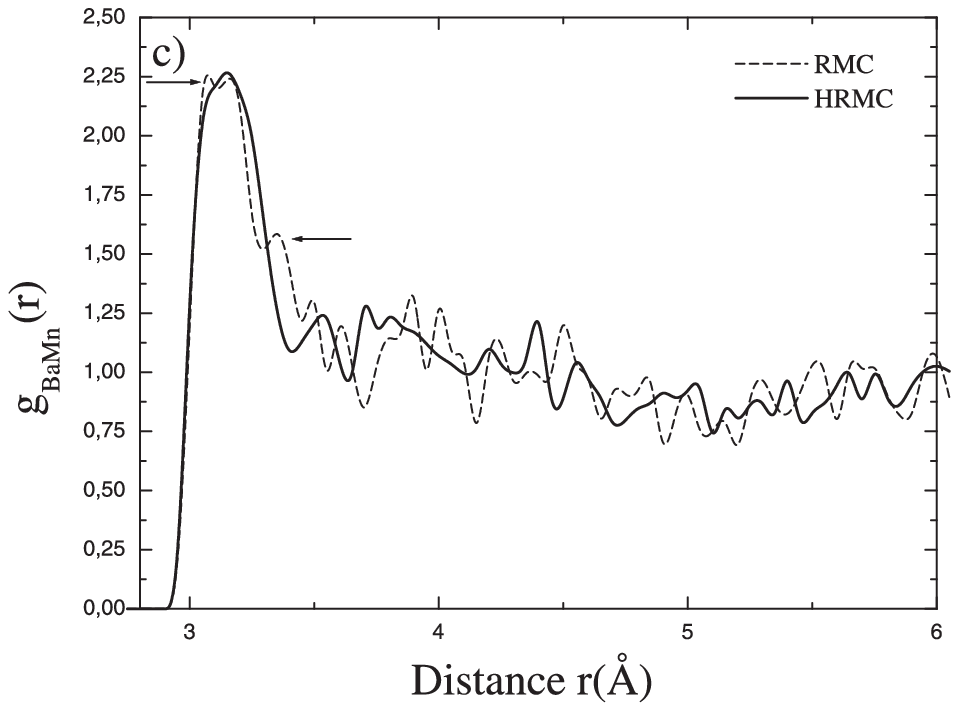}}
\caption{RMC and HRMC: partial pair distribution functions of $g_{\mathrm{BaFe}}(r)$ (a), $g_{\mathrm{FeMn}}(r)$ (b) and $g_{\mathrm{BaMn}}(r)$~(c). }
\label{fig3}
\end{figure}
One can make the same remarks concerning the other PDFs. As is clearly seen in figure~\ref{fig3}, it shows a meaningful improvement. Many artificial satellite peaks are alleviated in the first coordination of $g_{\mathrm{BaFe}}(r)$ [see arrows in figure~\ref{fig3}~(a)]. We also note that the first coordination of Fe--Mn figure~\ref{fig3}~(b) and Ba--Mn figure~\ref{fig3}~(c) is well marked by HRMC computation.

  The selected energy potential term plays an important role in alleviating the problem of the presence of unrealistic features. We can say that the energy penalty term is capable of providing a realistic description of the atomic interaction and helps to make a structural study for such a system.

\section{Conclusion}
\label{conc}
In the present work, we apply a hybrid reverse Monte Carlo method to the study of an additional energy constraint as a combined potential model between Coulomb and Lennard-Jones, in a fluoride glass system. Some drawbacks in RMC simulation were observed, such as the  artifacts that appear especially in PDFs. This can be due to the limited set of experimental data and/or due to the non-unique models of RMC. To solve this problem, we propose to incorporate the selected potential as an energy constraint in order to test its effectiveness. The results obtained by correlation functions show a good agreement between the method used  and the selected interaction model. Some artifacts that appeared in PDFs were eliminated by HRMC computation. Nevertheless, this study provides some important data used for a structural study of the BaMnMF7. As a final idea, it should be noted that the potential model used in this study can play an important role in describing the interactions between atoms, and in calculating the structural properties of fluoride glass or similar systems. One could also conclude that the HRMC method represents an essential tool for testing the interaction potential model, and may be used in conventional methods such as MD and MC simulation.

%
%% If you have problems with typesetting in ukrainian uncomment lines below.
%
%  \lastpage
%  \end{document}
\newpage
\ukrainianpart

\title{Вивчення застосовності об'єднаної потенціальної моделі до системи флюоридового
скла з використанням методу гібридного реверсного Монте Карло }

\author{С.М. Меслі\refaddr{label1,label2}, М. Габші\refaddr{label1,label3}, М. Котбі\refaddr{label2}, Г.~Ху\refaddr{label4}}

\addresses{
\addr{label1} Фізичний факультет, університет Хасіба Бен Буалі, Шлеф
02000, Алжир
\addr{label2} Фізичний факультет, університет A.B. Белкаіда , BP 119
Тлемсен 13000, Алжир
\addr{label3} Підготовча школа з природничих наук і техніки, Тлемсен
13000, Алжир
\addr{label4} Інститут фізики, університет Поля Верлена, 57000 Метц,
Франція
}

\makeukrtitle

\begin{abstract}
\tolerance=3000%
Вибір доречних моделей взаємодії є серед головних труднощів стандартних
методів, таких як молекулярна динаміка (МД) і Монте Карло (МК). З
іншого боку, так званий метод реверсного Монте Карло (РМК), що
ґрунтується на експериментальних даних, може бути застосований без
жодних міжатомних і/чи міжмолекулярних взаємодій. Результати РМК
супроводжуються  нереалістичними сателітними піками. Щоб уникнути цієї
проблеми, ми використовуємо розширення РМК алгоритму, яке вводить
додатковий енергетичний член у критерії аксептансу. Цей
метод називається методом гібридного реверсного Монте Карло (ГРМК).
Ідея цієї статті полягає у перевірці застосовності об'єднаної кулонівської
і леннард-джонсівської потенціальної моделі до системи флюоридового
скла BaMn$M$F$_{7}$ ($M={}$Fe,V) з використанням методу ГРМК.
Результати показують добре узгодження між експериментальними і
обчисленими характеристиками, а також значне покращення парціальних
парних функцій розподілу (ПФР). Ми припускаємо, що ця модель повинна
використовуватися в обчисленнях структурних властивостей і при описі
середніх кореляцій між компонентами флюоридового скла чи подібної
системи. Ми також припускаємо, що ГРМК могло б бути корисним для тестування  взаємодіючих потенціальних моделей, а також для
стандартних застосувань.
\keywords РМК моделювання, нереалістичні властивості, гібридне РМК
модлювання, потенціал Леннарда-Джонса, флюоридове скло

\end{abstract}


\begin{thebibliography}{99}
\bibitem{1}Zhang~J.-X., Li~H., Zhang~J., Song~X.-G., Bian~X.-F.,
    Chin. Phys. B, 2009, \textbf{18}, 4949; \doi{10.1088/1674-1056/18/11/055}.
\bibitem{2} Opletal~G., Petersen~T.C., O'Malley~B., Snook~I., McCulloch~D.G., Yarovsky~I.,
    Comput. Phys. Commun., 2008, \textbf{178}, 777; \doi{10.1016/j.cpc.2007.12.007}.
\bibitem{3}Dove~M.T., Tucker~M.G.,  Keen~D.A.,
    Eur. J. Mineral., 2002, \textbf{14}, 331; \doi{10.1127/0935-1221/2002/0014-0331}.
\bibitem{4} Frenkel~D., Smit~B., Understanding Molecular Simulations: from Algorithms to
Applications (2nd edition), Academic Press, San Diego, 2002.
\bibitem{5}Jorgensen~W.L., Chandrashekhar~J., Madura~J.D., Impey~R.W., Klein~M.L.,
    J. Chem. Phys., 1983, \textbf{79}, 926; \doi{10.1063/1.445869}.
\bibitem{6}McGreevy~R.L., Nucl. Instrum. Methods Phys. Res., Sect. A, 1995, \textbf{1}, 354; \doi{10.1016/0168-9002(94)00926-0}.
\bibitem{7}Evans~R.,
    Mol. Simul., 1990, \textbf{4}, 409; \doi{10.1080/08927029008022403}.
\bibitem{8}Pikunic~J., Clinard~C., Cohaut~N., Gubbins~K.E., Guet~J.M., Pellenq~R.J.-M., Rannou~I., Rouzaud~J. N.,
    Langmuir, 2003, \textbf{19}, 8565; \doi{10.1021/la034595y}.
\bibitem{9}Jain~K.S., Pellenq~R.J.-M., Pikunic~J.P., Gubbins~K.E.,
    Langmuir, 2006, \textbf{22}, 9942; \doi{10.1021/la053402z}.
\bibitem{10}Le~Bail~A., In: Proceedings of the Conference on Non-Crystalline Inorganic Materials: ``CONCIM-2003'' (Bonn, Germany,  2003), J. Non-Crys. Solids, Elsevier, 2005.
\bibitem{11}Habchi~M., Mesli~S.M., Kotbi~M., Xu~H.,
    Eur. Phys. J. B, 2012, \textbf{85}, 255; \doi{10.1140/epjb/e2012-21027-2}.
\bibitem{12}McGreevy~R.L.,
    J. Phys. Condens. Matter, 2001, \textbf{13}, R877; \doi{10.1088/0953-8984/13/46/201}.
\bibitem{13}Bartczak~W.M., Kroh~J., Zapalowski~M., Pernal~K.,
    Philos. Trans. R. Soc. London, 2001, \textbf{359}, 1785 \\ \doi{10.1098/rsta.2001.0867}.
\bibitem{14}McGreevy~R.L., Pusztai~L.,
    Mol. Simul., 1988, \textbf{1}, 359; \doi{10.1080/08927028808080958}.
\bibitem{15}Metropolis~N., Rosenbluth~A.W., Rosenbluth~M.N., Teller~A.H., Teller~E.,
    J. Chem. Phys., 1953, \textbf{21}, 1087; \\ \doi{10.1063/1.1699114}
\bibitem{16}Dove M.T., Tucker M.G., Wells A.S., Keen D.A., EMU Notes in
Mineralogy, 2002, \textbf{4}, 59--82.

\bibitem{17}Opletal~G., Petersen~T.C., O\'Malley~B., Snook~I.K., McCulloch~D.G., Marks~N., Yarovsky~I.,
    Mol. Simul., 2002, \textbf{28}, 927; \doi{10.1080/089270204000002584}.
\bibitem{18}Opletal~G., Petersen~T.C., McCulloch~D.G, Snook~I.K., Yarovsky~I.,
    J. Phys. Condens. Matter, 2005, \textbf{17}, 2605; \doi{10.1088/0953-8984/17/17/008}.
\bibitem{19}Putintsev~N.M., Putintsev~D.N.,
    Dokl. Phys. Chem., 2004, \textbf{399}, 278; \doi{10.1023/B:DOPC.0000048074.96169.7a}.
\bibitem{20}Kaplan~I.G.,
    Intermolecular Interactions: Physical Picture, Computational Methods and Model Potentials, John Wiley and Sons, Ltd., 2006.
\bibitem{21}Zhdanov~E.R., Fakhretdinov~I.A.,
J. Mol. Liq., 2005, \textbf{120}, 51; \doi{10.1016/j.molliq.2004.07.029}.
\bibitem{22}Beerdsen~E., Dubbeldam~D., Smit~B., Vlugt~T.J.H., Calero~S.,
    J. Phys. Chem. B, 2003, \textbf{107}, 12088; \\ \doi{10.1021/jp035229q}.
\bibitem{23}Hildebrand~F.E., Abeyaratne~R.,
    J. Mech. Phys. Solids, 2008, \textbf{56}, 1296; \doi{10.1016/j.jmps.2007.09.006}.
\bibitem{24}Vrabec~J., Stoll~J., Hasse~H.,
    J. Phys. Chem. B,  2001, \textbf{105}, 12126; \doi{10.1021/jp012542o}
\bibitem{25}Pusztai~L., Hars\'{a}nyi~I., Dominguez~H., Pizio~O.,
    Chem. Phys. Letts., 2008, \textbf{457}, 96; \doi{10.1016/j.cplett.2008.03.091}.
\bibitem{26}Kotbi~M., Xu~H., Habchi~M., Dembahri~Z.,
    Phys. Lett. A, 2003, \textbf{315}, 463; \doi{10.1016/S0375-9601(03)01014-4}.
\bibitem{27}Petersen~T.C., Yarovsky~I., Snook~I., McCulloch~D.G., Opletal~G.,
    Carbon, 2003, \textbf{41}, 2403; \\ \doi{10.1016/S0008-6223(03)00296-3}.
\bibitem{28}Steinczinger~Z., Pusztai~L.,
    Condens. Matter Phys., 2012, \textbf{15}, 23606; \doi{10.5488/CMP.15.23606}.

\end{thebibliography}
\end{document}